\documentclass[review]{elsarticle}

\usepackage{hyperref,amsmath,amssymb,graphicx}

\journal{Physics Letters A}

\biboptions{numbers,sort&compress}
\hypersetup{
	colorlinks=true,
	linkcolor=black
}

\begin{document}
	
	\begin{frontmatter}
		
		\title{Using Optical Systems to Simulate Topological Systems in Momentum Space and Measure Their Topological Numbers}
		
		 \author[mymainaddress,mysecondaddress]{Zhongcheng Feng}
		\ead{12231252@mail.sustech.edu.cn}
		
		\author[mymainaddress,mysecondaddress]{Jiansheng Wu\corref{mycorrespondingauthor}}
		\ead{wujs@sustech.edu.cn}
		
		\cortext[mycorrespondingauthor]{Corresponding author}
		
		\address[mymainaddress]{Shenzhen Institute for Quantum Science and Engineering, Southern University of Science and Technology, Shenzhen 518055, China.}
		\address[mysecondaddress]{International Quantum Academy, Shenzhen 518055, China.}
		
		\begin{abstract}
			We propose a new scheme for optical quantum simulation of topological systems: by using optical systems to simulate the variation of eigenstates of topological systems in momentum space, we can obtain the information of topological numbers. In this paper the scheme is applied to the one-dimensional (1D) Su-Schrieffer-Heeger (SSH) model and the two-dimensional (2D) Bernevig-Hughes-Zhang (BHZ) model. In addition, in order to apply our scheme to 2D topological systems, we design a method of calculating topological numbers by line integral.  Furthermore, we propose a more effective optical simulation scheme for the 2D topological system: we do the optical simulation around discontinuity points to obtain the vorticity of every discontinuity points and  the topological number is just the sum of the vorticity of all discontinuity points. 
		\end{abstract}
		
		
	\end{frontmatter}
	
	
	\section{Introduction}
	In recent years, using optical quantum systems to simulate topological systems has become a hot topic \cite{kitagawa2010exploring,asboth2012symmetries,asboth2013bulk,chalabi2019synthetic,zhan2017detecting,xiao2018higher,xiao2020non,chen2018observation}. The reason is that the optical system has the following advantages: horizontal polarization and vertical polarization of light can be used as two bases of particle states, and any polarization of light can be used to represent a particle state. Then the variation of light along the optical path can simulate the variation of particle states in the physical process \cite{kitagawa2010exploring}. In this way, optical systems can simulate topological systems with two-band states, such as those with two sites in the primitive cell, or a combination of electron state and hole state, or a combination of spin up state and spin down state. Then under time evolution, the quantum walk (QW) process of particle states can be simulated by the variation  of light along the optical path composed of the corresponding optical devices.
	Furthermore, the topological properties of the topological systems can be analyzed by observing the polarization distribution obtained after multi-step QWs. At present, there are two schemes to simulate topological systems with optical systems. One is the scheme to distinguish different positions on the spatial dimension \cite{zhan2017detecting,xiao2018higher,xiao2020non}, and the other is the scheme is  to distinguish different positions on the time dimension \cite{chen2018observation}.
	
	The above schemes for simulating topological systems with optical systems are both simulating the evolution of particle states in real space. We propose a new scheme: simulating the variation of the eigenstates of topological systems in momentum space with optical systems. Then by observing the variation of eigenstates in momentum space, we can obtain some topological properties defined on momentum space, such as topological numbers. 
	
	\section{Scheme of simulating topological systems with optical systems }
	For a topological system with two bands, the Hamiltonian in momentum space is a two-by-two matrix, which can be represented by the forms with Pauli matrices $\sigma_i$ $(i=x,y,z)$ \cite{sakurai1995modern}
	\begin{equation}
		H(k)=d_0(k)+d_x(k)\sigma_x+d_y(k)\sigma_y+d_z(k)\sigma_z,
	\end{equation}
	where $d_i(k)\ (i=0,x,y,z)$ are parameters with momentum $k$ as the variable.
	The corresponding  eigenvalues and eigenstates in general form can be solved as follows 
	\begin{equation}
		E_+=d_0+\sqrt{d_x^2+d_y^2+d_z^2},\ |\psi_+(k)\rangle=\begin{pmatrix}
			\cos\frac{\theta(k)}{2}\\
			\sin\frac{\theta(k)}{2}e^{i\phi(k)}
		\end{pmatrix},
		\label{equation 2}
	\end{equation}
	\begin{equation}
		E_-=d_0-\sqrt{d_x^2+d_y^2+d_z^2},\ |\psi_-(k)\rangle=\begin{pmatrix}
			\sin\frac{\theta(k)}{2}\\
			-\cos\frac{\theta(k)}{2}e^{i\phi(k)}
		\end{pmatrix},
		\label{equation 3}
	\end{equation}
	where $\theta(k)$ and $\phi(k)$  are defined as
	\begin{equation}
		\sin\theta(k)=\frac{\sqrt{d_x^2+d_y^2}}{\sqrt{d_x^2+d_y^2+d_z^2}},\ \cos\theta(k)=\frac{d_z}{\sqrt{d_x^2+d_y^2+d_z^2}},\label{equation 4}
	\end{equation}
	\begin{equation}
		\sin\phi(k)=\frac{d_y}{\sqrt{d_x^2+d_y^2}},\ 
		\cos\phi(k)=\frac{d_x}{\sqrt{d_x^2+d_y^2}}.
		\label{equation 5}
	\end{equation}

\begin{figure}[htbp]
		\centering
		\includegraphics[width=0.55\textwidth]{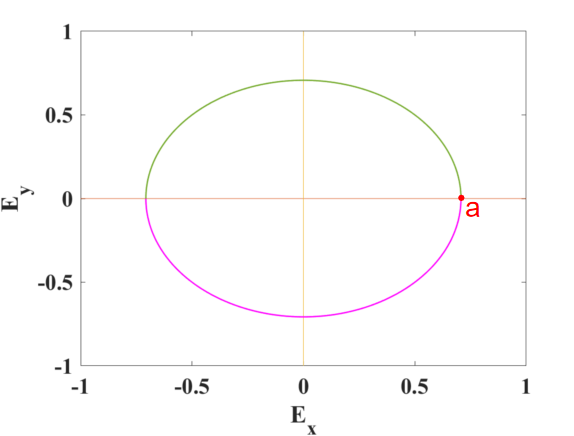}
		\caption{Polarization diagram of light when $\theta(k)=\pi$ and $\phi(k)=\pi/2$. \scriptsize{The pink line is obtained when $t\in [0,\frac{T}{2}]$ and the green line is obtained when $t\in[\frac{T}{2},T]$. Then the direction of the polarization is from point $a$ along the pink line, and then along the green line. So it is a left-handed polarization.}}
		\label{figure 1}
	\end{figure}

	We know the light has horizontal polarization and vertical polarization, which exactly correspond to two bases of the eigenstates in Eq. (\ref{equation 2}) and (\ref{equation 3}). In our simulations, the amplitudes of horizontal polarizations are  the first components of eigenstates, and the amplitudes of vertical polarizations are the norms of the second components of eigenstates, and the phase differences between vertical polarizations and horizontal polarizations are the phases of the second components of eigenstates. Thus the light corresponding to the eigenstates of two bands is
	\begin{equation}
		E_x^+(k)=\cos\frac{\theta(k)}{2}\cos(\omega t),\ E_y^+(k)=\sin\frac{\theta(k)}{2}\cos(\omega t+\phi(k)),
	\end{equation}
	\begin{equation}
		E_x^-(k)=\sin\frac{\theta(k)}{2}\cos(\omega t),\ E_y^-(k)=\cos\frac{\theta(k)}{2}\cos(\omega t+\phi(k)+\pi),
	\end{equation}
	where $\omega$ is the frequency of light. For unity of form, in the following $\phi(k)$ represents the phase differences between vertical polarizations and horizontal polarizations for two bands (which isn't above for negative-energy band). We get the polarization diagram of light when $t$ goes through one period $T=\frac{2\pi}{\omega}$. This can be measured by polarization state analyzer (PSA). Fig. \ref{figure 1} show the polarization diagram of light when $\theta(k)=\pi$ and $\phi(k)=\frac{\pi}{2}$. From Eq. (\ref{equation 4}) we get $\theta\in [0,\pi]$, and then the first components of eigenstates are nonnegative. So when $t=0$ the initial point is always in the first or fourth quadrant like point $a$ in fig. \ref{figure 1}. Thus fig. \ref{figure 1} show a left-handed polarization light. Moreover when $\phi(k)=0$ the polarization is a straight line in first and third quadrants, and when $\phi(k)=\pi$ or $-\pi$ the polarization is a  straight line in second and fourth quadrants, and when $\phi(k)\in (0,\pi)$ there are left-handed polarizations, and when $\phi(k)\in (-\pi,0)$ there are right-handed polarizations. So we can obtain the magnitude of $\phi(k)$ by the type of polarization. 
	
	In this way we simulate the eigenstates in momentum space with light. And the variation of eigenstates along a path (or boundary) in momentum space can also be simulated by a sequence of optical devices. For example, the variation of two neighboring eigenstates of the positive-energy band  can be represented by $U(k_{i+1},k_i)$
	\begin{gather}
		U(k_{i+1},k_i)=|\psi(k_{i+1})\rangle\langle\psi(k_i)| \label{equation 8}\\
		=\begin{bmatrix}
			1 & 0\\
			0 & e^{i\phi(k_{i+1})}
		\end{bmatrix}
		\begin{bmatrix}
			\cos\theta(k_{i+1})\cos\theta(k_{i}) & \cos\theta(k_{i+1})\sin\theta(k_{i})\\ 
			\sin\theta(k_{i+1})\cos\theta(k_{i}) & \sin\theta(k_{i+1})\sin\theta(k_{i}) 
		\end{bmatrix}
		\begin{bmatrix}
			1 & 0\\
			0 & e^{-i\phi(k_{i})}
		\end{bmatrix}. \notag
	\end{gather}
	We find that $U(k_{i+1},k_i)$ can be simulated with a ~`lens combination'. The first and third matrices in Eq. (\ref{equation 8}) can be simulated with  two phase-shifting lenses, and the second matrix can be simulated with a deflection lens converted between horizontal and vertical polarization. So a ~`lens combination' is composed of these three lenses.
	And the variation of any eigenstate and the initial state in momentum space can be expressed by the product of a series of variation of neighboring eigenstates:
	\begin{equation}
		U(k_{N},k_0)=U(k_{N},k_{N-1})\cdots U(k_{i+1},k_i)\cdots U(k_{1},k_0)
	\end{equation}
	So the process can be simulated with a series of 'lens combinations'.
	\begin{figure}[htbp]
		\centering
		\includegraphics[width=0.9\textwidth]{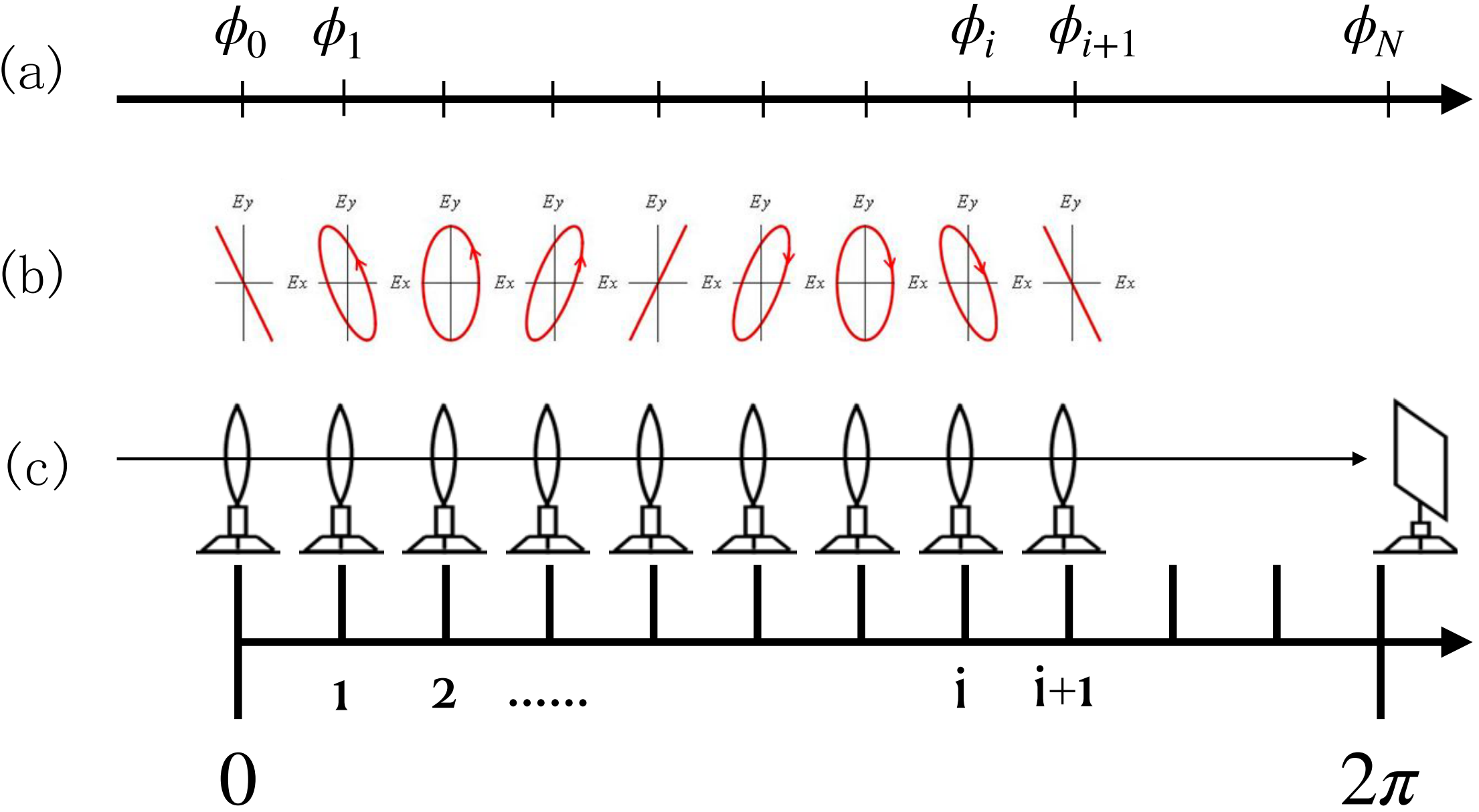}
		\caption{Schematic diagram of simulation devices. \scriptsize{(a) Phase differences of eigenstates at discrete points in momentum space. (b) Polarization diagrams of light corresponding to the eigenstates. (c) Schematic diagram of simulation devices. Each lens represents  ~`lens combination'.}}
		\label{figure 2}
	\end{figure}

	Fig. \ref{figure 2} is the schematic diagram of simulation devices. When light only goes through the ~`lens combination' at 0, we obtain  the polarization diagram of light corresponding to $|\psi(k_{0})\rangle$. And when light goes through the ~`lens combinations" at 0 and 1 we obtain  the polarization diagram of the light corresponding to $|\psi(k_{1})\rangle$, etc. Finally we obtain a series of polarization diagrams of light corresponding to eigenstates varying with momentum $k$. At this time, we observe the variation of the polarization of light with the variation of momentum $k$, then we can get the information of the topological number of the topological system.
	
	\section{Optical simulation of one-dimensional topological systems}
	
	The topological number of 1D topological system can be calculated by the formula \cite{li2015winding,guo2011topological}
	\begin{equation}
		c_n=\frac{1}{\pi i}\int\mathrm{d}k A_n(k),
	\end{equation}
	\begin{equation}
		A_n(k)=\langle \psi_n(k)|\partial_k|\psi_n(k)\rangle,
	\end{equation} 
	where $A_n(k)$ is Berry connection, $|\psi_n(k)\rangle$ ($n=+,-$) is the eigenstate in momentum space.
	We find that topological numbers of 1D topological systems can be approximately obtained by a series of eigenstates in momentum space. So we can try to use our scheme to obtain topological numbers.
	
	\subsection{One-dimensional SSH model}
	The most classical 1D topological model is the 1D Su-Schrieffer-Heeger (SSH) model \cite{su1979solitons}.
	Its Hamiltonian  in momentum space is denoted as
	\begin{equation}
		H(k)=\begin{bmatrix}
			0 & t_1+t_2e^{-ik}\\
			t_1+t_2e^{ik} & 0
		\end{bmatrix}
		=d_x\sigma_x+d_y\sigma_y,
	\end{equation}
	where $d_x=t_1+t_2\cos k$, $d_y=t_2\sin k$.
	According to Eq. (\ref{equation 2}) and (\ref{equation 3}), the eigenenergys and eigenstates are
	\begin{equation}
		E_+=\sqrt{d_x^2+d_y^2},\ |\psi_+(k)\rangle=\frac{1}{\sqrt{2}}
		\begin{pmatrix}
			1\\
			e^{i\phi(k)}
		\end{pmatrix};
	\end{equation}
	\begin{equation}
		E_-=-\sqrt{d_x^2+d_y^2},\ |\psi_-(k)\rangle=\frac{1}{\sqrt{2}}
		\begin{pmatrix}
			1\\
			-e^{i\phi(k)}
		\end{pmatrix}.
	\end{equation}
	Then we can compute the Berry connections of two bands
	\begin{equation}
		A_+=A_-=\frac{i}{2}\frac{\partial\phi(k)}{\partial k}.
	\end{equation}
	And we get the topological number
	\begin{equation}
		c_{\pm}=\frac{1}{2\pi}\int_{0}^{2\pi}\frac{\partial\phi(k)}{\partial k}\mathrm{d}k.
	\end{equation}
	We can find that topological number of the 1D SSH model is the accumulation of the phase difference between the second and the first components of the eigenstate divided by $2\pi$. 
	Therefore, in our scheme, we can obtain the topological number by observing the accumulation of the phase difference between the horizontal and vertical polarization of light.
	
	Through the topological band theory \cite{ashcroft2022solid,bansil2016colloquium,shen2018topological}, we can obtain the topological number of the two cases: 
	when $t_1>t_2$, $c_{\pm}=0$;
	when $t_1<t_2$, $c_{\pm}=1$. And $t_1=t_2$ is the phase transition point.
	
	\subsection{Optical simulation}
\begin{figure}[htbp]
		\centering
		\includegraphics[width=0.9\textwidth]{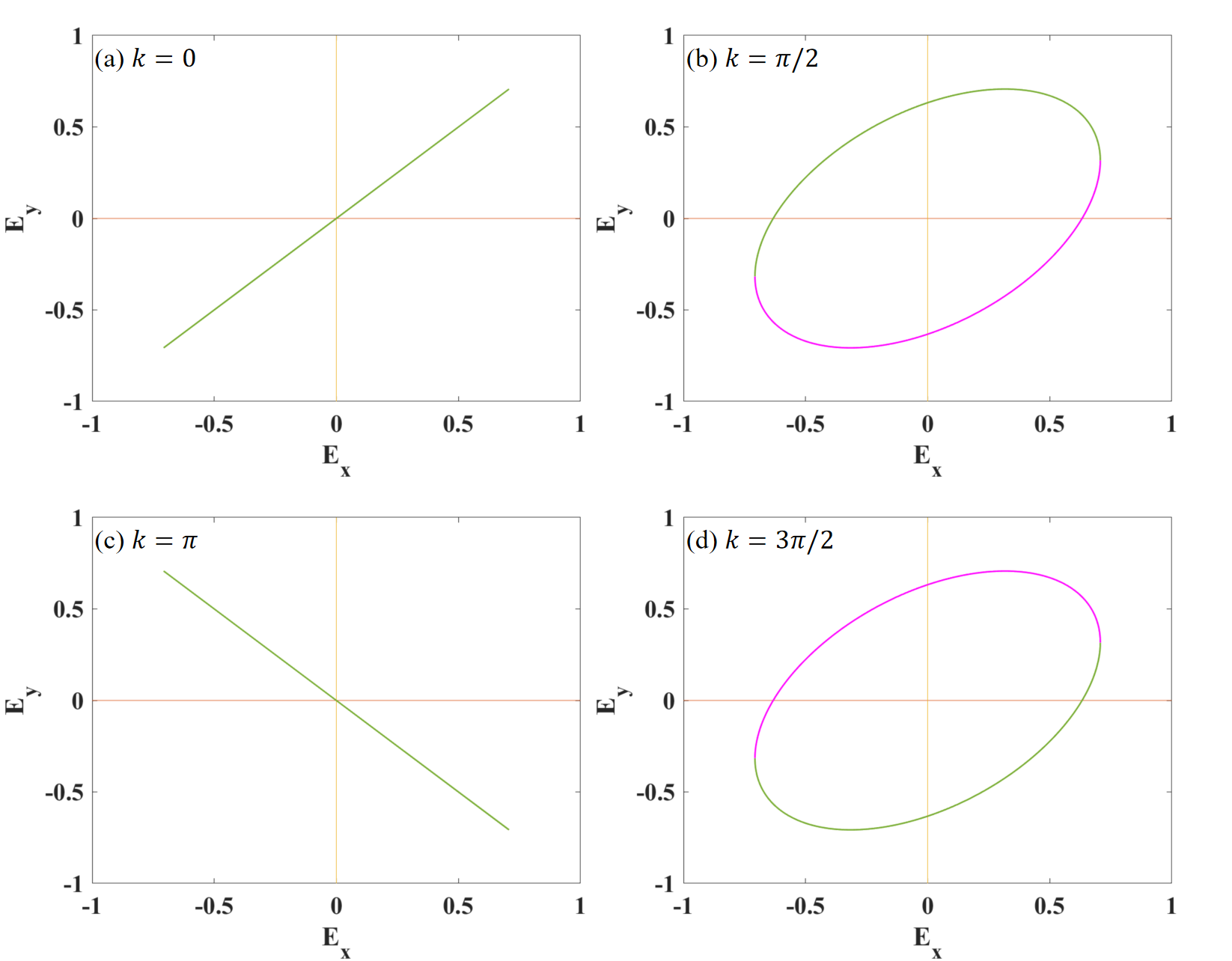}
		\caption{Polarization diagrams when $t_1=1$, $t_2=2$. \scriptsize{(a)-(d) The polarization diagrams of light corresponding to the positive-energy eigenstates $|\psi_+(k)\rangle$ when $k$ is 0, $\pi/2$, $\pi$ and $3\pi/2$ respectively (When $k=2\pi$ the polarization of light return to that when $k=0$). We can see that when k varies from 0 to $2\pi$, the phase difference $\phi(k)$ varies from 0 to $\pi$ through left-handed polarization and then to $2\pi$ through right-handed polarization. So the accumulation of phase difference is $2\pi$ when $t_1=1$, $t_2=2$.}}
		\label{figure 3}
	\end{figure}

\begin{figure}[htbp]
		\centering
		\includegraphics[width=0.9\textwidth]{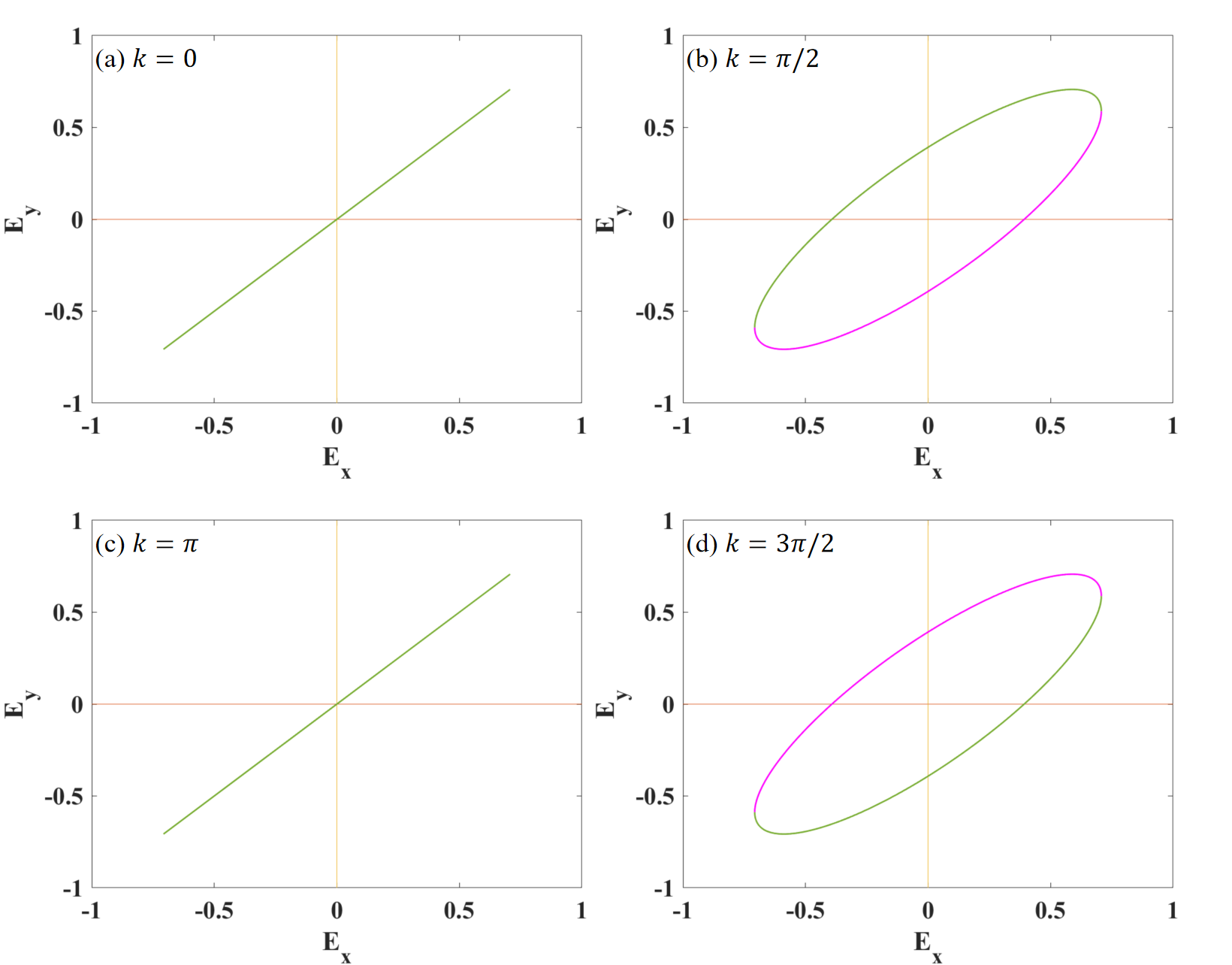}
		\caption{polarization diagrams when $t_1=3$, $t_2=2$. \scriptsize{(a)-(d) The polarization diagrams of light corresponding to the positive-energy eigenstates $|\psi_+(k)\rangle$ when $k$ is 0, $\pi/2$, $\pi$ and $3\pi/2$ respectively. When k varies from 0 to $2\pi$, the phase difference $\phi(k)$ varies from 0 to 0 through left-handed polarization and again to 0 through right-handed polarization. So the accumulation of phase difference is 0 when $t_1=3$, $t_2=2$.}}
		\label{figure 4}
	\end{figure}

	Now we try to find how to get this result by optical simulation. Fig. \ref{figure 3} shows the polarization diagrams of light corresponding to the positive-energy eigenstates $|\psi_+(k)\rangle$ when $t_1=1$, $t_2=2$. We obtain the accumulation of phase difference is $2\pi$. So the topological number of the positive-energy band is $1$ in this case.
	And fig. \ref{figure 4} shows the polarization diagrams of light corresponding to the positive-energy band $|\psi_+(k)\rangle$ when $t_1=3$, $t_2=2$. We obtain the accumulation of phase difference is $0$. So the topological number of the positive-energy band is 0 in this case.	

	We find that the results of two simulations agree well with the results of  theory. Therefore, we realize the purpose of simulating the variation of eigenstates in momentum space with optical system in 1D topological system to obtain the topological number by observing the variation of polarization of light. Next, we further try to apply our scheme to 2D topological systems.
	
	\section{Optical simulation of two-dimensional topological systems}
	
	The  formula for the topological number of the 2D topological system \cite{thouless1982quantized,avron1983homotopy,kohmoto1985topological,niu1985quantized} is
	\begin{gather}
		c_n=\frac{1}{2\pi i}\int_{T^2}\mathrm{d}^2kF_{12}(k),\\
		F_{12}=\partial_1A_2(k)-\partial_2A_1(k),
	\end{gather}
	where $T^2$ is the Brillouin zone, $F_{12}$ is Berry curvature and $A_{\mu}(k)$ is Berry connection. The topological number of a 2D topological system is the surface integral of the Berry curvature in the Brillouin zone.
	Because our scheme can only simulate the variation of eigenstates along a path in momentum space, we need to change the method of calculating topological number from surface integral to line integral. So we design a method to compute the topological numbers of 2D topological systems by line integral.
	
	\subsection{Method of calculating topological numbers by line integral}
	\subsubsection{Conditions of calculation}
	
	By using the Stokes' theorem, we can convert the surface integral of the Berry curvature into the line integral of the Berry connection \cite{shen2012topological,bernevig2013topological}
	\begin{equation}
		c_n=\frac{1}{2\pi i}\oint_{\partial T^2}A_{\mu}(k)\mathrm{d}k_{\mu}.
		\label{equation 22}
	\end{equation}
	However it is valid only if eigenstates are continuous and smooth over the entire Brillouin zone. For topologically nontrivial  systems the topological number is exactly an obstruction to the continuity of eigenstates in the whole Brillouin zone, and there must be discontinuity points  in the Brillouin zone \cite{bernevig2013topological}. Therefore, we cannot simply use the Stokes theorem to convert the surface integral of Berry curvature into the line integral of Berry connection. Obviously we need to set certain conditions to use the Stokes' theorem.
	
	For every eigenstates in momentum space, we can put any global phases to the eigenstates, and the newly obtained states are still eigenstates. The process of adding global phases to eigenstates in momentum space to obtain new eigenstates is called gauge transformation. By gauge transformation we get a gauge for eigenstates with specific phase \cite{xiao2010berry}. Under a specific gauge, we find all discontinuity points in momentum space. Then line integrals can be used to calculate topological numbers only if they satisfy the two conditions:\\
	$\it Condition$ 1: We need to design new boundaries (i.e. the path of the integrals) , which include all discontinuity points of the eigenstates, such that the eigenstates inside the boundaries are continuous and smooth. \\
	$\it Condition$ 2: Since eigenstates at the discontinuity points are not well defined (the phase defined can be arbitrary), we need to replace eigenstates at the discontinuity points by eigenstates at momentum adjacent to discontinuity points and inside the boundaries to obtain the phase accumulation.
	
	When numerically calculating the topological number, we discretize the momentum along the boundary and adopt the following approximation \cite{fukui2005chern}
	\begin{eqnarray}
		\ln\langle \psi_n(k)| \psi_n(k+\delta k_\mu)\rangle\approx\ln(1+\langle \psi_n(k)|\partial_\mu| \psi_n(k)\rangle\delta k_\mu)\nonumber\\
		\approx\langle \psi_n(k)|\partial_\mu| \psi_n(k)\rangle\delta k_\mu= A_\mu(k)\delta k_\mu=\tilde{A}_\mu(k),
	\end{eqnarray}
	where $\tilde{A}_\mu(k)$ the line integral unit for Berry connection.
	Assuming that the discrete points on the boundary $C$ is $k_l$, the topological number can be expressed as
	\begin{equation}
		c_n=\frac{1}{2\pi i}\sum_{k_l\in C}\tilde{A}_\mu(k_l).
	\end{equation}
	
	\subsubsection{Two-dimensional BHZ model}
	
	We use Bernevig-Hughes-Zhang (BHZ) model \cite{bernevig2006quantum} to verify the feasibility of this method.
	The Hamiltonian of the 2D BHZ model in momentum space is denoted by
	\begin{equation}
		H=(m-2t_s(\cos k_x+\cos k_y))\sigma_z+t_{so}\sin k_x\sigma_x+t_{so}\sin k_y\sigma_y.
	\end{equation}
	We have $d_x=t_{so}\sin k_x$, $d_y=t_{so}\sin k_y$, $d_z=m-2t_s(\cos k_x+\cos k_y)$. We obtain eigenenergys and eigenstates as Eq. (\ref{equation 3}) and (\ref{equation 4}).
	When $t_s=1$, $t_{so}=2$, the topological numbers calculated by topological band theory \cite{ashcroft2022solid,bansil2016colloquium,shen2018topological} are:
	when $|m|>4t_s$, $c_{\pm}=0$;
	when $0<m<4t_s$, $c_+=-1, c_-=1$;
	when $-4t_s<m<0$, $c_+=1,c_-=-1$.
	And $m=0,\pm4t_s$ are the phase transition points.
	
	Now let us use the method of calculating topological numbers by line integral to get topological numbers. 
	Firstly, we define a gauge that is to completely define the eigenstates of the topological system.
	We choose the gauge such that the first components of eigenstates are positive real numbers. So we need to remove the phases of the first components from the resulting eigenstates (having arbitrary global pheses) by globally multiplying by the inverse of phases of the first components
	\begin{equation}
		e^{if(k)}=|||\psi_n(k)\rangle_1||/|\psi_n(k)\rangle_1,\ 
		|\tilde{\psi}_n(k)\rangle=e^{if(k)}|\psi_n(k)\rangle.
	\end{equation}
	Under this gauge, we get eigenstates as Eq. (\ref{equation 2}) and (\ref{equation 3}). And discontinuity points are points at which the first components of the eigenstates are zero, making those eigenstates are not well defined (global phases of eigenstates are not fixed) under this gauge. At $(0,0)$, $(0,\pi)$, $(\pi,0)$ and $(\pi,\pi)$, $d_x=d_y=0$ such that $\theta=0$ or $\pi$, and thus 
	$\sin\frac{\theta}{2}=0$ or $\cos\frac{\theta}{2}=0$ making the first componets of eigenstates are zeroes. So all discontinuity points of eigenstates of two bands are $(0,0)$, $(0,\pi)$, $(\pi,0)$ and $(\pi,\pi)$.
	
\begin{figure}[htbp]
		\centering
		\includegraphics[width=0.9\textwidth]{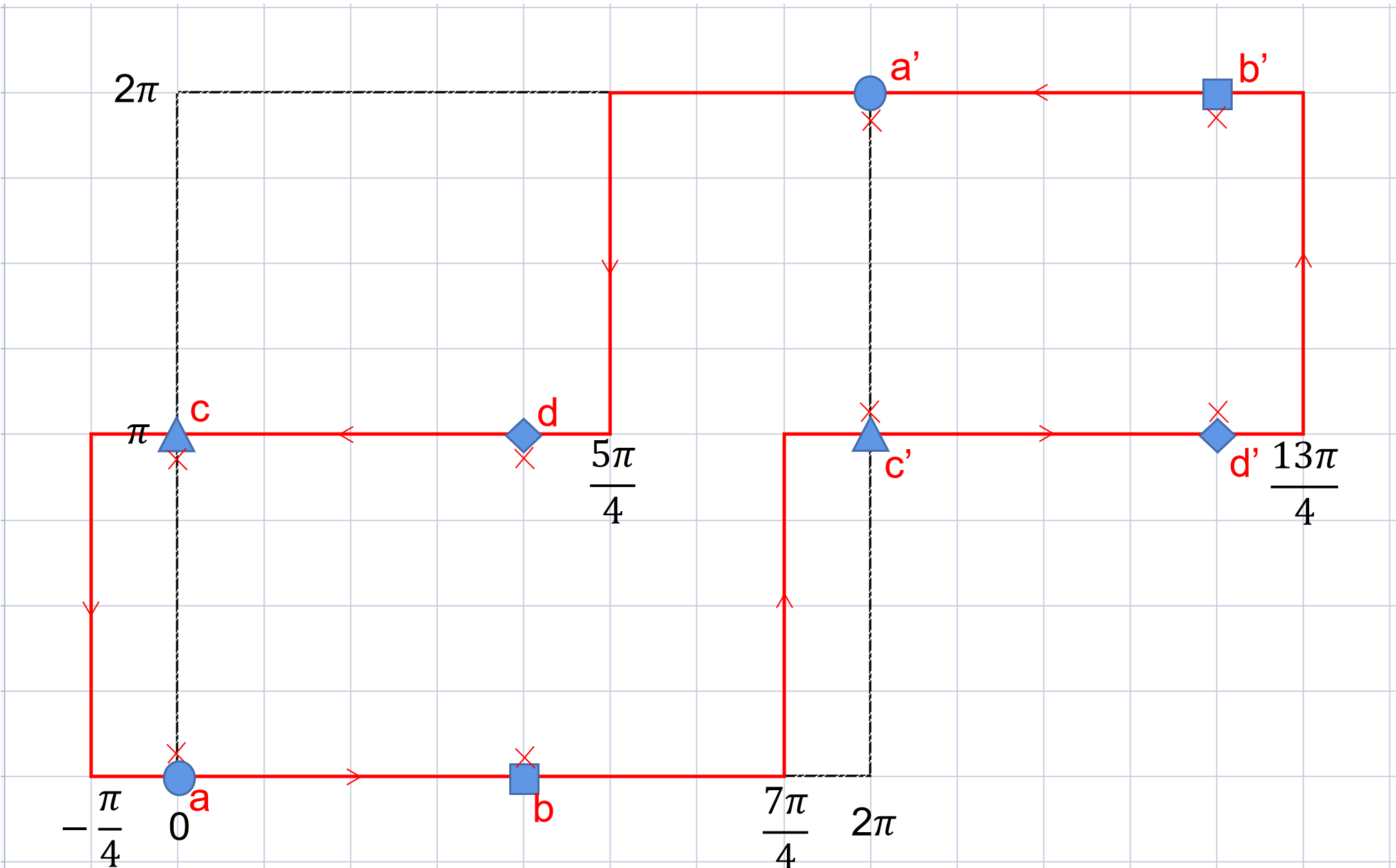}
		\caption{Integral boundary. \scriptsize{The eigenstates of the cross points replace the eigenstates of discontinuity points.}}
		\label{figure 5}
	\end{figure}

	According to $Condition$ 1 (boundary places all discontinuity points), the integral boundary we designed is shown in fig. \ref{figure 5}. At the same time, the eigenstate of each discontinuity point is represented by the eigenstate of the cross point (as illustrated in the  fig. \ref{figure 5}) to satisfy $Condition$ 2.

	The final result of our numerical calculation is shown in table \ref{table 1} when $t_s=1$, $t_{so}=2$.
	As shown in table \ref{table 1}, our results are consistent with the topological numbers obtained by the topological band theory. It shows that our method can calculate topological numbers of 2D topological systems.
	\begin{table}[htbp]
		\begin{center}
			\begin{tabular}{c|c|c|c|c|c|c|c|c}
				\hline
				m & -7 &-5 & -3 & -1 & 1 & 3 & 5 & 7\\
				\hline
				$c_+$ & 0 & 0 & 0.9998 & 0.9997 & -0.9997 & -0.9998 & 0 & 0\\
				\hline
				$c_-$ & 0 & 0 & -0.9998 & -0.9997 & 0.9997 & 0.9998 & 0 & 0\\
				\hline
			\end{tabular}
		\end{center}
		\caption{Topological numbers when $t_s=1$, $t_{so}=2$. \scriptsize{The distance between neighbouring discrete points is $\pi/40$ , and the distance between discontinuity points and cross points is $10^{-3}\pi$.}}
		\label{table 1}
	\end{table}

	\subsection{Optical simulations along the boundary}
\begin{figure}[htbp]
		\centering
		\includegraphics[width=0.9\textwidth]{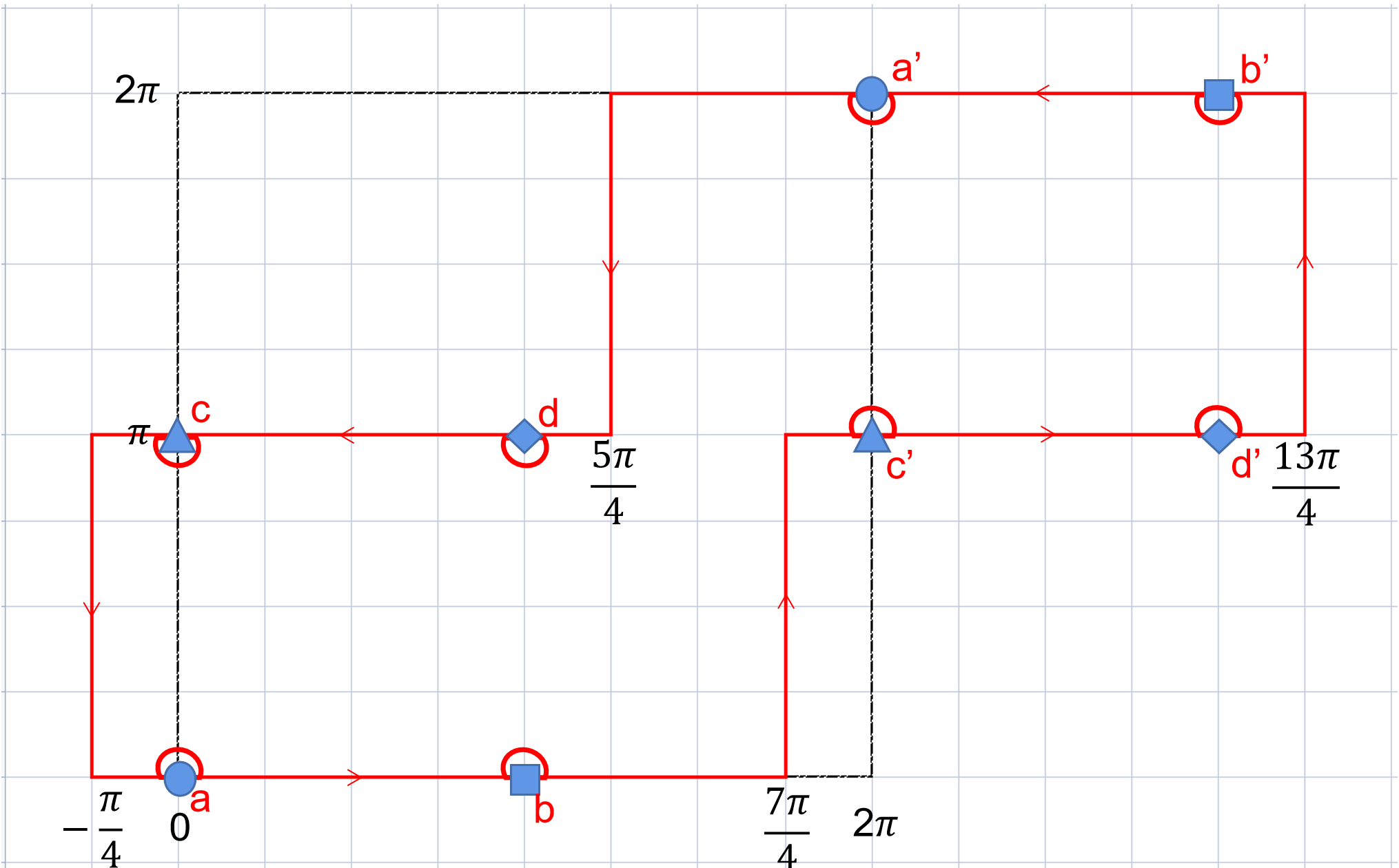}
		\caption{Optical simulation along the boundary. \scriptsize{There are semicircles around discontinuity points inside the boundary to take discrete points, making  phase differences continuous along the boundary.}}
		\label{figure 6}
	\end{figure}

	In the previous subsection, we design a method to calculate the topological numbers of 2D topological systems by line integral, and then we can adopt our optical simulation scheme to do optical simulation along the boundary shown in fig. \ref{figure 5}.	However it's not enough to only take one point around each discontinuity point. It causes discontinuous phase differences between eigenstates adjacent to discontinuity points inside the boundary and their neighboring eigenstates. So we take more points around discontinuity points and eventually form small semicircles around the discontinuity points  as shown in fig. \ref{figure 6}.

	In addition we find that except eigenstates around discontinuity points, eigenstates of the continuous parts of the boundary always have equivalent eigenstates at different positions on the boundary. For example as shown in fig. \ref{figure 6}, the eigenstates between $a$ and $b$ are the same as the eigenstates between $a'$ and $b'$. And when we do optical simulation, their  simulation directions are just opposite such that their phase accumulation gotten by optical simulation are cancelled each other. Then the phase accumulation left is  optical information about eigenstates around discontinuity points. At this point we can combine the simulations at every pairs of equivalent discontinuity points ($a$ and $a'$, $b$ and $b'$, $c$ and $c'$, $d$ and $d'$) to form simulations of respective closed loops. In fact the accumulation of phase differences clockwise around a discontinuity point divided by $2\pi$ is called the vorticity at that discontinuity point \cite{bernevig2013topological}. And the topological number is the sum of vorticity at all discontinuity points.
	\begin{equation}
		n=\frac{1}{2\pi}\int_{\partial(R_s^\epsilon)}\mathrm{d}k\cdot \nabla\chi(k)
	\end{equation}
	where $\chi(k)$ is the phase difference between the first and second components of eigenstates $\chi(k)=-\phi(k)$, and $R_s^\epsilon$ is a small region around each discontinuity point. 
	So we find a more effective optical simulation for 2D topological systems: we perform the optical simulation clockwise around all discontinuities to obtain the vorticity of each discontinuity point, and finally sum up vorticity of all discontinuity points to obtain the topological number.
	
	\subsection{Optical simulations around discontinuity points}
	\subsubsection{The type of discontinuous points}
	
\begin{figure}[htbp]
		\centering
		\includegraphics[width=0.9\textwidth]{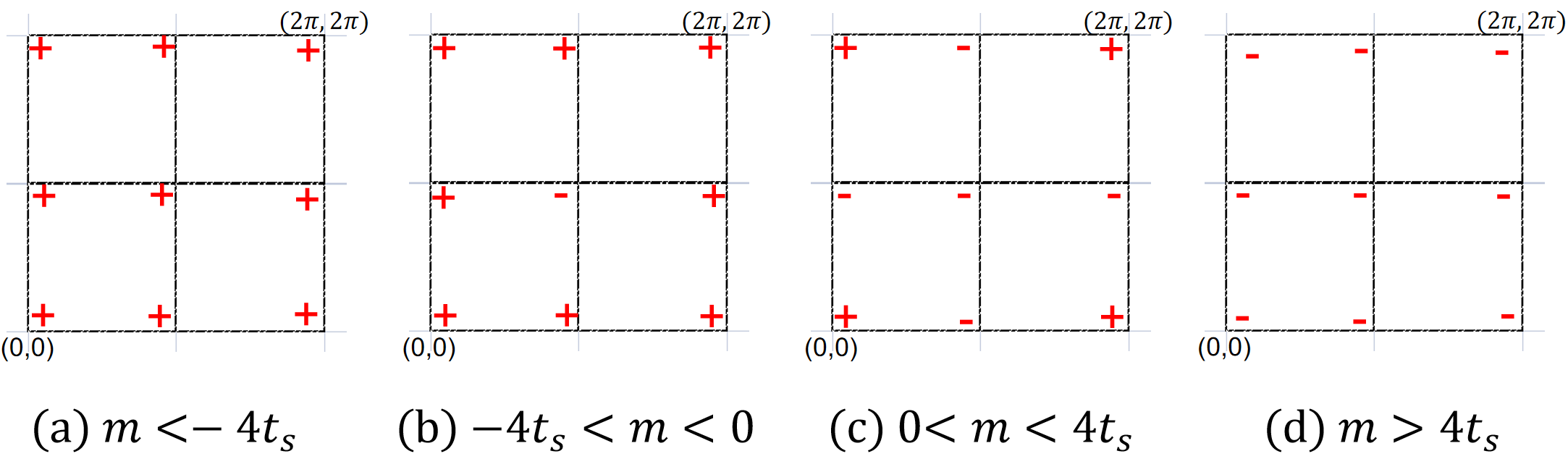}
		\caption{ Discontinunity points under different conditions. \scriptsize{'+' indicates that here the discontinuity point belongs to the positive-energy band, and '-' indicates that here the discontinuity point belongs to the negative-energy band.} }
		\label{figure 7}
	\end{figure}

	When the parameter $m$ is in different intervals, discontinuity points belong to different energy bands. From Eq. (\ref{equation 4}) we get that at four discontinuity points: when $d_z>0$, $\theta=0$ and then $\cos\frac{\theta}{2}=1$, $\sin\frac{\theta}{2}=0$, the first components of the negative-energy eigenstates are zero so that those discontinuity points belong to the negative-energy band; when $d_z<0$, $\theta=\pi$ and then $\cos\frac{\theta}{2}=0$, $\sin\frac{\theta}{2}=1$, the first components of the positive-energy eigenstates are zero so that those discontinuity points belong to the positive-energy band.
	By comparing the magnitude of m and $t_s$, we can get the sign of $d_z$:
	when $m<-4t_s$, at all discontinuity points $d_z<0$;
	when $-4t_s<m<0$, at $(0,0)$, $(0,\pi)$, $(\pi,0)$ $d_z<0$ and at $(\pi,\pi)$ $d_z>0$;
	when $0<m<4t_s$, at $(0,0)$ $d_z<0$ and at $(0,\pi)$, $(\pi,0)$, $(\pi,\pi)$ $d_z>0$;
	when $m>4t_s$, at all discontinuity points $d_z>0$.
	Thus we get the types of all discontinuity points under different conditions of $m$ as shown in fig. \ref{figure 7}.

	\subsubsection{Optical simulations around discontinuity points}

\begin{figure}[htbp]
		\centering
		\includegraphics[width=0.9\textwidth]{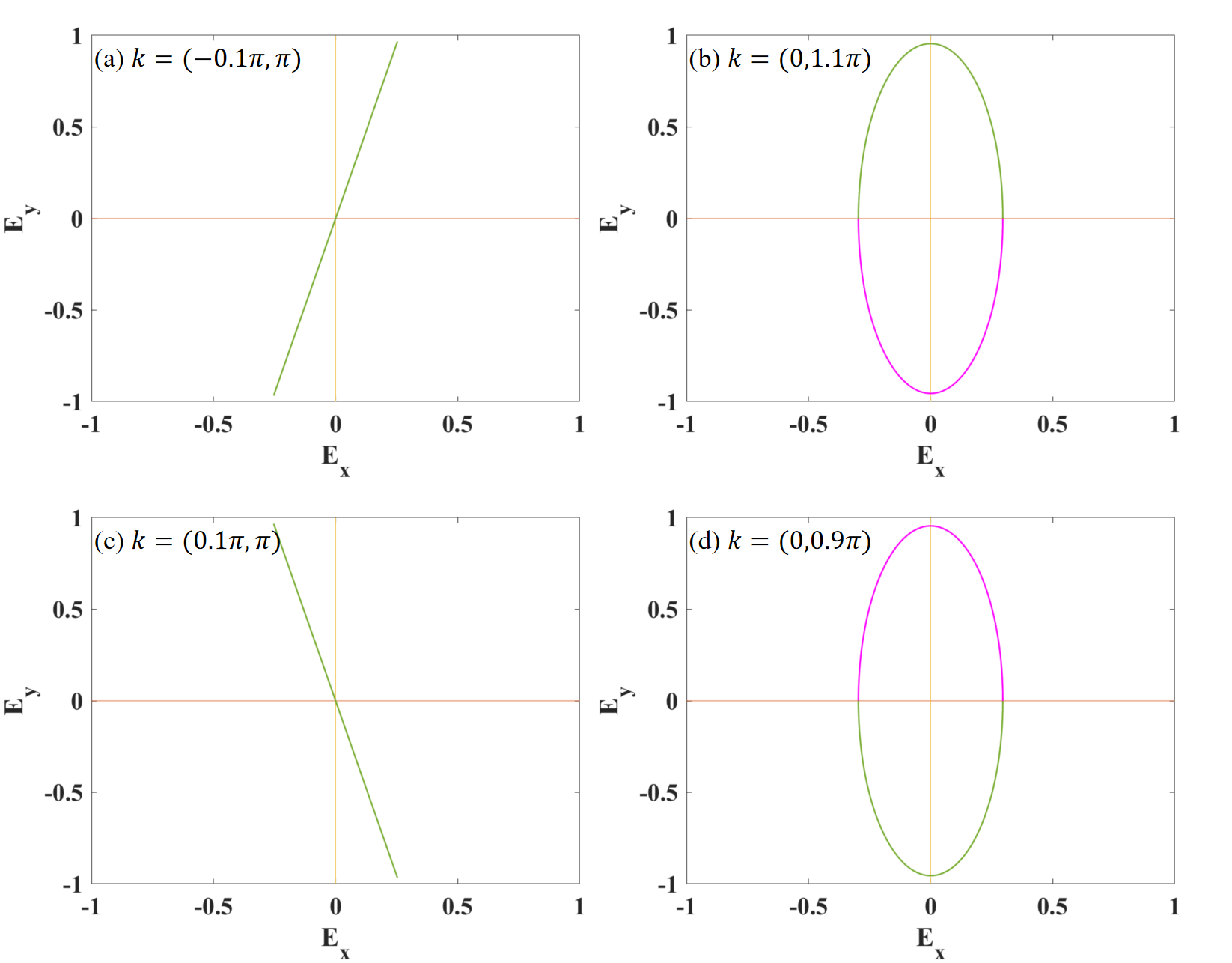}
		\caption{Polarization diagrams of light corresponding to negative-energy eigenstates near $(0,\pi)$ when $m=1$, $t_s=1$, $t_{so}=2$.  \scriptsize{(a)-(d) {The polarization diagrams of eigenstate with momentum close to and on the left, upper, right, and lower of  $(0,\pi)$}. When k varies clockwise around $(0,\pi)$, the phase difference $\phi(k)$ varies from 0 to $\pi$ through left-handed polarization and then to $2\pi$ through right-handed polarization. So the accumulation of phase difference clockwise around $(0,\pi)$ is $2\pi$.}}
		\label{figure 8}
	\end{figure}

\begin{figure}[htbp]
		\centering
		\includegraphics[width=0.9\textwidth]{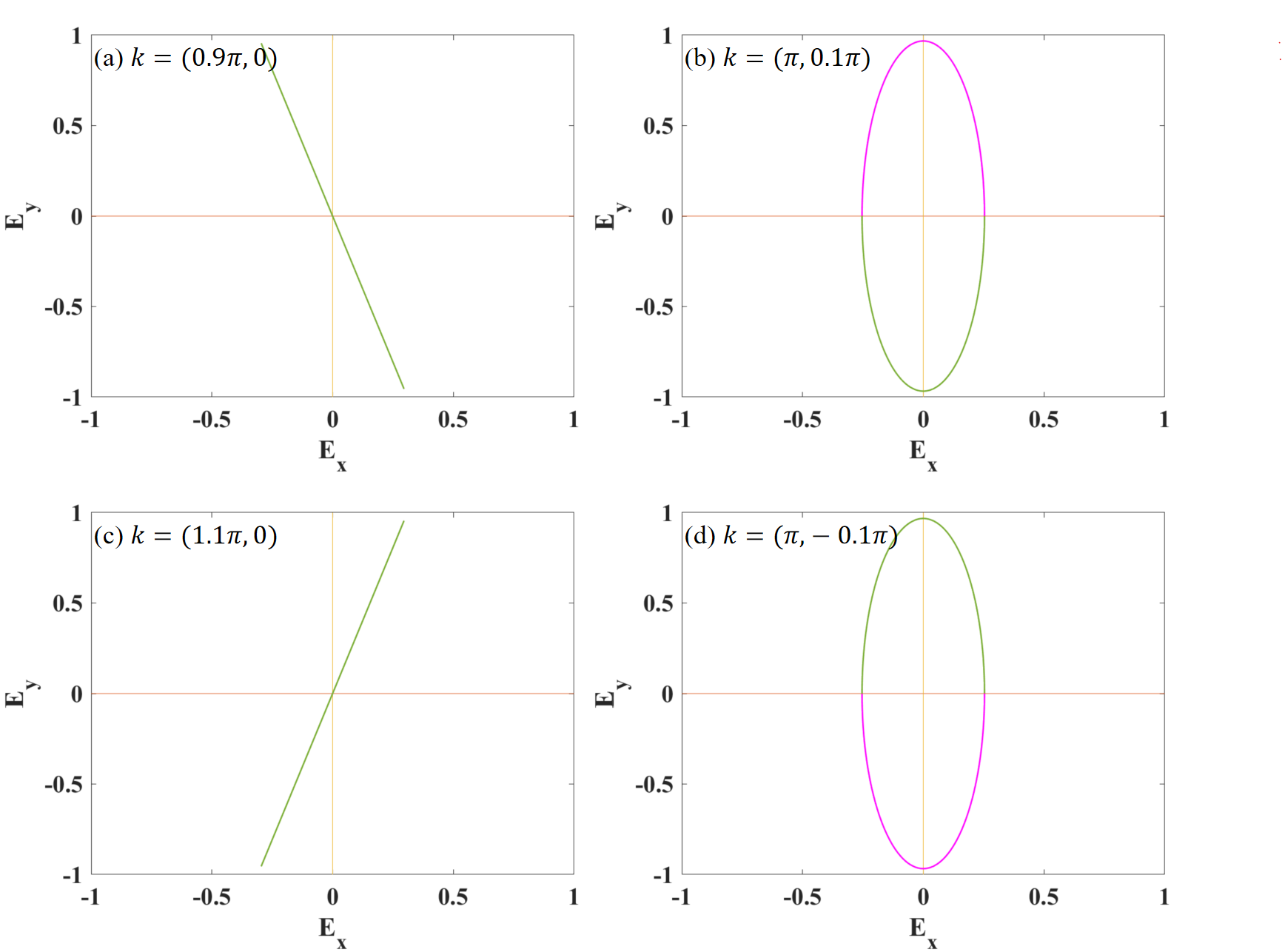}
		\caption{Polarization diagrams of light corresponding to negative-energy eigenstates near $(\pi,0)$ when $m=1$, $t_s=1$, $t_{so}=2$.  \scriptsize{(a)-(d) The polarization diagrams of eigenstate with momentum close to and on the left, upper, right, and lower of $(\pi,0)$. When k varies clockwise around $(\pi,0)$, the phase difference $\phi(k)$ varies from $-\pi$ to 0 through right-handed polarization and then to $pi$ through left-handed polarization. So the accumulation of phase difference clockwise around $(\pi,0)$ is $2\pi$.}}
		\label{figure 9}
	\end{figure}

\begin{figure}[htbp]
		\centering
		\includegraphics[width=0.9\textwidth]{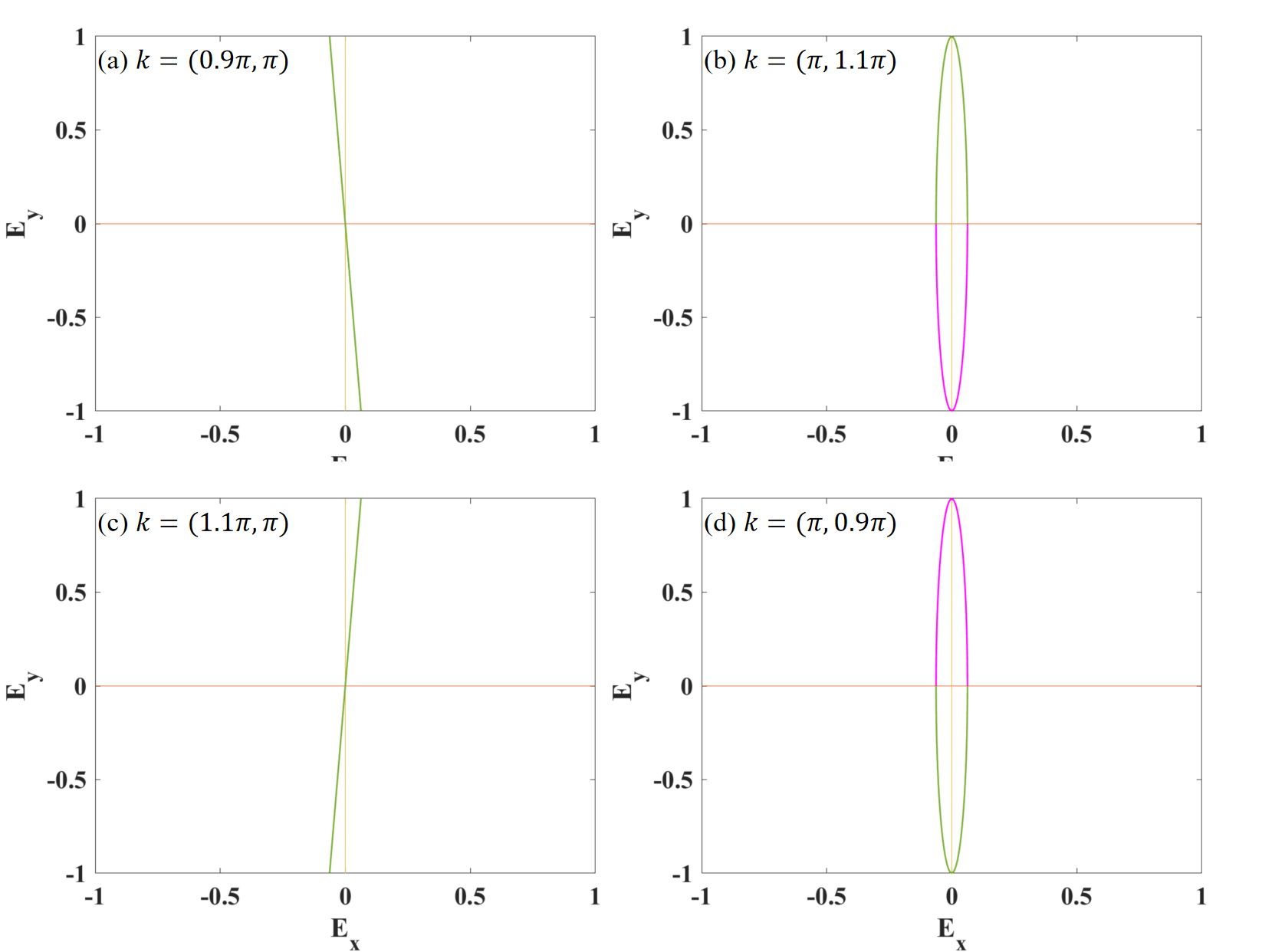}
		\caption{Polarization diagrams of light corresponding to negative-energy eigenstates near $(\pi,\pi)$ when $m=1$, $t_s=1$, $t_{so}=2$.  \scriptsize{(a)-(d)  The polarization diagrams of eigenstate with momentum close to and on the left, upper, right, and lower of  $(\pi,\pi)$. When k varies clockwise around $(\pi,0)$, the phase difference $\phi(k)$ varies from $\pi$ to $0$ through left-handed polarization and then to $-pi$ through right-handed polarization. So the accumulation of phase difference clockwise around $(\pi,\pi)$ is $-2\pi$.}}
		\label{figure 10}
	\end{figure}

	Now we start to perform optical simulations clockwise around discontinuity points.
	For example, for the two-dimensional BHZ model, when $m=1$, $t_s=1$, $t_{so}=2$, the discontinuity points $(0,\pi)$, $(\pi,0)$ and $(\pi,\pi)$ belong to the negative-energy band, and the topological number of the negative-energy band is $1$. The polarization diagrams of light corresponding to negative-energy eigenstates near $(0,\pi)$, $(\pi,0)$ and $(\pi,\pi)$ when $m=1$, $t_s=1$, $t_{so}=2$ are shown in fig. \ref{figure 8}-\ref{figure 10}. We get the accumulation of phase difference clockwise around $(0,\pi)$, $(\pi,0)$ is $2\pi$, and clockwise around $(\pi,\pi)$ is $2\pi$. Then the vorticity at $(0,\pi)$, $(\pi,0)$ is 1 and at $(0,\pi)$ is $-1$. Thus when $m=1$, $t_s=1$, $t_{so}=2$, the topological number of the negative-energy band, the sum of vorticity at all discontinuity points, is 1, which is the same as that obtained by the topological band theory. Similarly, optical simulations can be done for both positive-energy and negative-energy bands under other conditions, and topological numbers under those conditions also can be obtained.

	\section{Conclusion}
	
	In this paper, we introduce a scheme of simulating topological systems with optical systems: we simulate the variation of eigenstates in momentum space by optical systems and obtain the information of topological numbers by measuring the phase accumulation. We apply the scheme to the 1D SSH model and 2D BHZ model. In the 1D  SSH model, we observe the accumulation of the phase difference between the horizontal polarization and the vertical polarization of light along the 1D momentum space to obtain the topological number. In the 2D BHZ model, we need to find a 1D boundary (path) inside which the eigenstates are continuous. Then the topological number can be obtain by line integral of Berry connection along this boundary. Furthermore, we find a more effective simulation scheme: we do optical simulation around the discontinuity points to get the vorticity of each discontinuity point, and sum the vorticity of all discontinuity points to get the topological number.

	\bibliography{document}
		
\end{document}